\def\tit#1{``#1,''}
\long\def\comment#1{}

\newcommand{\ket}[1]{| #1 \rangle}

\newcommand\eq[1]{Eq.~(\ref{#1})}

\documentclass{article}

\renewenvironment{abstract}%%
{\noindent \hskip.35in\begin{minipage}{5.75in} \parindent.2in%
   \noindent \normalsize \bf Abstract: \rm}{\hfil \end{minipage}}

\begin{document}
\title{\bf Universal quantum processors \\
       with arbitrary radix \boldmath{$n \ge 2$}}
\author{{\bf Alexander Yu.\ Vlasov}\\
\footnotesize\em FRC/IRH, Mira Street 8, 197101, St.--Petersburg, RUSSIA}
\date{March 2001}
%\sloppy
\maketitle
\begin{abstract}
 Here is discussed the Hamiltonian approach to construction of
deterministic universal (in approximate sense) programmable quantum
circuits with qubits or any other quantum systems with dimension
of Hilbert space is $n \ge 2$.
\end{abstract}

\noindent
Let us suggest, that universal quantum processor can be described as
circuit with three ``buses'': quantum, intermediate, and pseudo-classical.
It is discussed with more details elsewhere (in \cite{Vlasov2001} based
on results about quantum control \cite{Jozsa95} and programmability
\cite{Chuang97}). Here are mentioned only topics related with theme of
present paper about universality.

If we have finite set of universal (in approximate sense \cite{Cleve99})
quantum gates, we may approximate any unitary operator with necessary
precision \cite{Cleve99,Deutsch85,Deutsch89,Ekert95,Gates95,DiVincenzo95,%
Lloyd96,Vlasov2000a,Vlasov2000b}. There exist sets of two-gates both for
binary quantum circuits with qubits
\cite{Cleve99,Gates95,DiVincenzo95,Lloyd96,Vlasov2000a} and for non-binary
ones with equal dimensions $n>2$ of Hilbert space of each elementary system
(sometime it is called qu$n$it) \cite{Vlasov2000b}.

Here is considering application of universal set of quantum gates in
three-level design of quantum processor described above. Intermediate bus
together with quantum data bus is example of quantum control
\cite{Jozsa95,Vlasov2001}, i.e., if state of intermediate bus is $\ket{p}$
then $p$-th gate $U_p$ from given set is applied to quantum data bus,
$\ket{d} \mapsto U_p\ket{d}$.

An approximation of given unitary operator $U$ is decomposition on product
of universal gates $U_k$:
\begin{equation}
 U = U_{p_l}\cdots U_{p_2} U_{p_1}
\label{decomp}
\end{equation}
And so it is necessary to apply quantum control $l$-times with value
of intermediate bus is $\ket{p_l}$ on $l$-th step \cite{Vlasov2001}.

For generation of the sequence of numbers $p_l$ may be used \cite{Vlasov2001}
reversible\footnote{It can be shown, that using irreversible classical gates
on intermediate and pseudo-classical buses together with inevitable problem
with quantum description of composite system with three buses may also cause
non-unitary evolution of quantum data bus} classical (in terminology used by
\cite{Ekert95}) circuits acting on pseudo-classical bus together with
intermediate bus. Due to this property of quantum processor it may be
important sometime to have well defined algorithm of decomposition
\eq{decomp} for arbitrary unitary operator $U$. Otherwise we should keep all
$l$ numbers as table and then size of pseudo-classical bus is proportional to
{\em maximally possible} length $l_{\mathrm{max}}$ of simulation (``cyclic
Q-ROM approach'' \cite{Vlasov2001}).

An approach with clear algorithm of decomposition uses structure of Lie
algebra of unitary group \cite{Ekert95,DiVincenzo95,Lloyd96,%
Vlasov2000a,Vlasov2000b} and from physical point of view related with
Hamiltonian of quantum circuits. In such approach main object is not a
unitary matrix $U$, but Hamiltonian of the evolution, i.e.  Hermitian matrix
$H$ with property
\begin{equation}
 U(\tau) = e^{i\,H\,\tau},\quad \exists t,\ U = U(t)
\label{expH}
\end{equation}
for some real parameters $t, \tau$, time.

The exponent \eq{expH} makes possible for small $\tau$ to use approximation
with sums instead of products. Say if Hermitian matrix $H$ can be
decomposed on some basis $H_I$ in space of Hermitian matrices, then
\begin{equation}
 U(\tau) = \exp(i\sum_I {n_I H_I \tau}) \approx \prod_I{\exp(i\,n_I H_I \tau)},
\quad \tau \to 0.
\label{Utau}
\end{equation}
Let $U_I = \exp(i\, H_I \tau)$ and $n_I$ may be approximated by natural
number (integral part) if $\tau$ is small and $n_I$ are big. Then:
\begin{equation}
 \exp(i\sum_I {n_I H_I \tau}) \approx \prod_I{U_I^{n_I}},
 \quad U \equiv U(t) \approx \bigl(\prod_I{U_I^{n_I}}\bigr)^{t/\tau}
\label{Ut}
\end{equation}
where $t/\tau$ may be considered as natural number or approximated
by it for small $\tau$.

But number of elements of basis of Hermitian matrix for quantum circuit is
exponentially large. Say for system with $k$ qubits it is $4^k$ and it
is $n^{2k}$ for other radix $n$. Due to such property, as Hamiltonians of
universal quantum gates are chosen only small subset with property, that
any other element of basis can be generated by using sequence of commutators
\begin{equation}
 H_{[JK]} \equiv i (H_J H_K - H_K H_J)
\end{equation}
To produce $U_{[JK]} = \exp(i \, H_{[JK]} \tau)$ it is possible to
use approximation:
\begin{equation}
U_{[JK]} \approx U^{s_\tau}_J U_K^{s_\tau}U^{-s_\tau}_J U_K^{-s_\tau},
\quad s_\tau = 1/\sqrt{\tau}
\label{UJK}
\end{equation}
there for small $\tau$ $s_\tau$ again can be considered as a natural number
(integral part). Here small indexes like $k$ are used for set of initial
universal gates, $U_1$ and capital indexes like $J$ are compound,
$U_{[[12]3]}$.

Here is a problem with negative power, $-s_\tau$. It can be resolved
for qubit by special design \cite{Vlasov2000a} where for any universal gate
$U_k(t_0)$ is unit for same $t_0$ and so $U_k^{-s} = U_k^{t_0-s}$ with positive
$t_0-s$. For non-binary universal gates it is not necessary so, but it is
possible simply to double amount of universal gates and have $U_k^{-1}$
together with any gate $U_k$.

\medskip

Let us describe now algorithm of approximation in general. It is chosen some
small interval of time $\tau$, the less it the higher precision. For
approximation of some gate $U$ with Hamiltonian $H$ as in \eq{expH} may be
found linear decomposition of the $H$ by basis $H_I$ \eq{Utau}. Each
component with compound indexes $U_I$ is represented by universal gates $U_k$
as it was shown in \eq{UJK}, and finally approximation of $U$ is result of
nested cycles described by \eq{Utau}, \eq{Ut}.

\smallskip

The example shows, that approximation may demand many operations and number
of steps quickly grows with refinement of precision. But if there is some
method to implement structure of the approximating algorithm as reversible
classical circuit, then size of program register (defined mainly by length of
pseudo-classical bus) may be more appropriate. Due to it algebraic structures
similar with introduced in \cite{Vlasov2000a,Vlasov2000b} may be useful.

There is also important note about reversibility of algorithm of
approximation. Formally it is possible to use some irreversible circuit and
apply standard technique, i.e. instead of irreversible function
$f\colon a \mapsto f(a)$ to use reversible one with property
$F \colon (a,0) \mapsto (a,f(a))$, but then each step of algorithm will
produce new portion of ``junk'' and pseudo-classical bus again must
have size proportional to $l_{\mathrm{max}}$, and such a case maybe even
worst, than reversible cyclic Q-ROM register without any programming.

But even if such reversible algorithm is found, the property of universality
may be rather formal. It is clear from consideration above, that number
of steps is proportional to amount of non-vanishing terms $H_I$ in
decomposition of Hamiltonian. So, if problem area is related with absolutely
arbitrary unitary operators, then particular set of gates is not very
essential and length of simulation is exponential on number of qubits.

Of course such situation would not be realistic if number of qubits is big
enough and so set of basic universal gates should take into account
particular set of possible problems. Say it may be task to simulate any
possible quantum circuits composed by many different $k$-gates, with $k$ is
not very large, i.e. ``local'' quantum circuits.

\end{document}